\documentclass{PoS}

\title{A free parametrized TOV: Modified Gravity from Newtonian to Relativistic Stars}

\ShortTitle{A free parametrized TOV}

\author{\speaker{Hermano Velten}\thanks{This contribution was supported by A*MIDEX project (n ANR-11-IDEX-0001-02) funded by 
the ``Investissements d'avenir" French Government program, managed by the French National Research Agency (ANR). 
HV and AO also thank CNPq and FAPES. AW acknowledges support from INFN Sez. di Napoli (Iniziative Specifiche TEONGRAV). Moreover,
HV and AW would like to thank the Organizers for warm hospitality during the conference. We also thank David Edwin Alvarez Castillo for a critical reading of the manuscript. }\\
        CPT, Aix Marseille Universit\'e, UMR 7332, 13288 Marseille, France\\  
        Universidade Federal do Esp\'{\i}rito Santo (UFES), Vit\'oria, Brazil\\
        E-mail: \email{velten@pq.cnpq.br}}

\author{Adriano M. Oliveira\\
        Instituto Federal do Esp\'irito Santo (IFES), Guarapari, Brazil\\
				Universidade Federal do Esp\'{\i}rito Santo (UFES), Vit\'oria, Brazil\\
        E-mail: \email{adriano.oliveira@ifes.edu.br}}

\author{Aneta Wojnar\\
        IFT, University of Wroclaw, pl. M. Borna 9, 50-204, Wroclaw, Poland,\\
        INFN Sez. di Napoli, Univ. di Monte S. Angelo, Ed. G, 
        Via Cinthia, I-80126 Napoli, Italy.\\
        E-mail: \email{aneta.wojnar@ift.uni.wroc.pl}}

\abstract{We test a free {\it ad hoc} parametrization of the Tolman-Oppenheimer-Volkoff (TOV) equation. We do not have in mind any 
specific extended theory of gravity (ETG) but each new parameter introduced has a physical interpretation. Our aim is fully pedagogical 
rather than a proposal for a new ETG. Given a realistic neutron star equation of state we map the contributions of each new parameter into a shift
in trajectories of the mass-radius diagram. This exercise allows us to make the correspondence between each TOV sector with possible modifications
of gravity and clarifies how neutron star observations are helpful for distinguishing theories.}

\FullConference{The Modern Physics of Compact Stars 2015\\
		30 September 2015 - 3 October 2015\\
		Yerevan, Armenia}

\begin{document}

\section{Introduction}

One of the main challenges in modern astrophysics concerns the equation of state (EoS) of neutron stars (NSs). 
Recent observations point out to $M \sim 2 M_{\odot}$ \cite{MassNs, stellarcollapse} objects preferring therefore {\it stiff} nuclear EoS. 
However, it is worth noting that this latter conclusion is valid only in the general relativity (GR) domain.

Another important property of NSs is the stellar radius $R$. 
The mass-radius $(M-R)$ diagram has become a comprehensive tool for matching theoretical predictions to observations. 
Modifications of GR can produce shifts in the trajectories along the $M-R$ diagram but sometimes higher masses are 
achieved at the price of having undesirable larger radii configurations. 
The current radius determination is not so precise as in the mass case but future satellites will place very precise constraints on both $M$ and $R$ for the same object. As specific examples one can cite The Neutron star Interior Composition Explorer (NICER) mission and the SKA radio project.

Although GR is a very well tested theory and its predictions have been confirmed with solar system and binary pulsars observations the cosmological arena still challenges it. 
Actually, by adopting GR we are left with the strong indication that $\sim 95\%$ of the energy budget of the universe is 
unknown. 
According to the most recent observations, this fraction is divided into the probably fraction of $\sim25\%$ for dark matter and $\sim70\%$ for dark energy.

One possible solution relies in modifying GR on large (cosmological) scales. 
By replacing either the dark matter or the dark energy component by some extended theory of gravity (ETG) the already confirmed prediction of GR should be reconquered. 
This happens usually via the inclusion of screening mechanisms around an astrophysical environment where the ETG predictions should suffer a metamorphopsia back to the GR ones. 
If this is actually the case, it is therefore expected that the NS interior and its habitat remains described by GR. 
However, the use of NSs for constraining ETGs has become a fruitful route of investigation. 
See \cite{MGTNS} for a few recent works on this topic.

The first usual procedure is to compute the analogue of the Tolman-Oppenheimer-Volkoff (TOV) equation 
\cite{OV, Tolman:1939dn}---the GR structure for static and spherical objects---for your favorite ETG. 
In general, departures from the standard TOV equation are non-trivial. 
However, it can be useful to understand how masses and radii of NSs are shifted in the $M-R$ plane according to specific changes in the TOV structure.
In this sense, a recent and interesting work \cite{PostTOV} proposed a ``Post-TOV'' formalism based on the parametrized post-Newtonian theory in which non-GR effects can be separately studied in relativistic stars. 
We also point out the analysis done in Ref. \cite{Schwab:2008ce} in which the specific pressure contribution to the active gravitational mass in NSs is studied in detail. 

Our aim in this contribution is to propose a free parametrization for the TOV equation. 
The curious reader can go directly to Eqs. (\ref{eq:pTOV}) and (\ref{dMdrsigma}). 
In some sense, this is an extension of \cite{Schwab:2008ce}. 
Our approach should be seen as a pedagogical guide to understand the role played by different physical contributions to the TOV equation rather than new a ETG. 

In the next section we review a quick derivation of the TOV equation. 
Section \ref{secPTOV} presents our free parametrization for stellar equilibrium equations. 
We will adopt one realistic EoS for the NS interior and show in section \ref{secN} results on the $M-R$ plane when varying such new parameters. 
In the final section we discuss in more detail the role played by each parameter. We use $c=\hslash =1$ units.

\section{The Tolman-Oppenheimer-Volkoff equation}

In this section we review the derivation of the TOV equation which can be found in most textbooks about GR or relativistic astrophysics.

The first and simplest step in studying stellar objects is to consider that matter is spherically symmetric distributed in a static geometry i.e., time reversal about any origin of time,   
\begin{eqnarray}
ds^2 =  -B(r)dt^2+A(r)dr^2+r^2(d\theta^2+\sin^2\theta d\phi^2)\quad.
\label{eq:metrica}
\end{eqnarray}
where hereafter $B(r)\equiv B; A(r)\equiv A$. General relativity is based on Einstein's equations
\begin{eqnarray}
R_{\mu\nu} - \frac{1}{2}g_{\mu\nu} R = 8\pi G T_{\mu\nu}\quad,
\label{eq:Einstein}
\end{eqnarray}
where $R_{\mu\nu}$ is the Ricci tensor and $R$ is the curvature scalar (trace of tensor $R_{\mu\nu}$). 
The matter content, in this case, the internal description of the star, is encapsulated in the energy-momentum tensor  
\begin{eqnarray}
T_{\mu\nu} &=& (\rho + p) u_\mu u_\nu + p g_{\mu\nu}\quad,
\end{eqnarray}
where respectively $\rho$ and $p$ are the density and pressure of the fluid which depend on the radial coordinate only. Also, $u_\mu$ is the 4-velocity (with $u_\mu u^{\mu} = -1$). Noticed that the fluid is at rest, then $u_{r}=u_{\theta}=u_{\phi}=0$ and $u_t=-(-g^{tt})^{-1/2}=-\sqrt{B}$. 

In order to derive the TOV equation we need to consider the conservation $T^{\mu\nu}_{\quad;\nu}=0,$ which is concomitant with Bianchi's identities. This reads
\begin{equation}\label{equilib}
\frac{1}{B}\frac{d B}{dr}=-\frac{2}{\rho+p}\frac{d p}{dr}\quad.
\end{equation}

The information from the components $G_{tt}$, $G_{rr}$ and $G_{\theta \theta}$ provides the equation
\begin{equation}
\frac{d}{dr}\left( \frac{r}{A}\right)=1-8\pi G \rho r^2\quad.
\end{equation}
Its solution, demanding that $A(0)$ is finite, is
\begin{equation}
M(r)\equiv \int_0^{r}4\pi {r^{\prime}}^{2}\rho(r^{\prime}) dr^{\prime}\quad.
\label{eq:massa}
\end{equation}
This solution is similar to the Newtonian equation and therefore one calls $ M(r)$ the mass function for a given radius $r$. 
Adopting this association with the classical framework, the gravitational mass inside the star will be calculated using $M = M(R)$ where $R$ is the finite stellar radius.

Finally, it is also possible to combine all these equations into the final form   
\begin{eqnarray}
\frac{dp}{dr} &=& -\frac{G M(r) \rho}{r^2}\frac{\left(1+\frac{p}{\rho}\right)\left(1+\frac{4\pi r^3 p}{M(r)}\right)}{1 - \frac{2GM(r)}{r}}\quad,
\label{eq:TOV}
\end{eqnarray}
which is know as the TOV equation \cite{OV, Tolman:1939dn}.

\section{Parametrized TOV} \label{secPTOV}

The TOV equation (\ref{eq:TOV}) represents the full general relativistic equilibrium configuration for stars. 
The compactness $\eta = 2 G M /R$ of a star measures the relevance of GR effects. 
The Newtonian counterpart is enough for low compactness $\eta \ll 1$ stars. 
For instance, in white dwarfs $\eta_{WD} \sim 10^{-6}$ whereas in main sequence stars $\eta_{MS} \sim 10^{-4}$. 
Typical NS compactness values are in the range $0.2\lesssim\eta_{NS}\lesssim 0.4$. In general, relativistic effects can be captured by pressure contributions and curvature effects. 
Also, having in mind that typical predictions of ETG replace the gravitational coupling by some effective quantity we therefore propose for the equilibrium equation the following parametrization
\begin{eqnarray}
\frac{dp}{dr} &=& -\frac{G(1+\alpha) \mathcal{M}(r) \rho}{r^2}\frac{\left(1+\frac{\beta p}{\rho}\right)\left(1+\frac{\chi 4\pi r^3 p}{\mathcal{M}(r)}\right)}{1 - \frac{\gamma 2G\mathcal{M}(r)}{r}}\quad.
\label{eq:pTOV}
\end{eqnarray}
We also generalize the mass function $\mathcal{M}(r)$ in (\ref{eq:pTOV}) by writing  
\begin{equation}
\frac{d \mathcal{M}(r)}{dr} = 4 \pi {r}^{2} (\rho + \sigma p)\quad.
\label{dMdrsigma}
\end{equation}
Note that this is an effective mass which is used in the integration of (\ref{eq:pTOV}). For $\sigma \neq 0$ this definition should be different from the conventional mass as calculated in (\ref{eq:massa}). However, the actual gravitational mass remains being $M\equiv M(R)$. Since this definition can also be written in terms of the metric components (as usually defined via $A(r)= e^{\lambda}$) there exists a alternative visualization for that which reads 
\begin{equation}
 \mathcal{M}(r)=r(1-e^{-\lambda(r)})\quad.
\end{equation}

Now, there are 5 new parameters, namely $\alpha$, $\beta$, $\gamma$, $\chi$ and $\sigma$.

\begin{itemize}
\item $\alpha$ parametrizes the effective gravitational coupling, i.e., $G_{eff}= G (1+\alpha)$. 
In particular, in $f(R)$ theories one has $\alpha=1/3$ \cite{brax}. In GR $\alpha=0$.
\item $\beta$ couples to the inertial pressure. 
The term $(\rho+p)$ appears from the hydrostatic equilibrium $T^{\mu \nu}_{\,\,\,\,\,; \nu = r}=0$, where $r$ is the radial coordinate. 
In GR $\beta=1$.
\item $\gamma$ is an intrinsic curvature contribution which is absent in the Newtonian physics, i.e., in the classical case $\gamma=0$. 
In GR $\gamma=1$. 
\item $\chi$ measures the active gravitational effects of pressure which is a remarkable feature of GR. 
Its effect has already investigated in Ref. \cite{Schwab:2008ce}. 
In GR $\chi=1$.
\item $\sigma$ changes the way the mass function is computed taking into account possible gravitational effects of pressure. In GR $\sigma=0$.
\end{itemize}

The strategy which will be adopted hereafter is to assume one realistic EoS and then vary such parameters. 
This somehow follows the reasoning discussed in Ref. \cite{Yavuz}. 
In this reference, it is argued that theoretical EoS used for NSs are only an order of magnitude larger than typical values in nucleon scattering experiments. 
On the other hand, curvature effects within NSs are many orders of magnitude above the domain where GR is confined.
Therefore, it is safer to claim we have a better knowledgment about the NS EoS than the gravitational theory inside these objects.

Some particular configurations of these parameters have the following interpretation:
\begin{itemize}
\item $\alpha=\beta=\gamma=\chi=\sigma=0$: This corresponds to the Newtonian hydrostatic equilibrium which gives rise to the Lane-Endem equation. 
\item $\alpha=0;\,\beta=1;\,\gamma=\chi=0;\,\sigma=3$: The neo-Newtonian hydrodynamic---a proposal which tries to include relativistic inspired pressure effects at Newtonian level---has been applied to the stellar equilibrium problem in Ref. \cite{Oliveira:2014jk}. 
Contrary to the Newtonian case, the neo-Newtonian case leads to the existence of maximum masses for NS as well as the GR prediction. 
However, the maximum masses found in the neo-Newtonian formalism are slightly higher than the GR ones.
\item $\alpha\neq 0;\,\beta=0;\gamma=\chi=\sigma=0$: This would consist in the simplest manifestation of modified gravity in Newtonian stars. 
In general, some $f(R)$ theories gives rise to the modification $\alpha=1/3$ together with $\sigma\neq0$. 
\item $\alpha=\beta=\gamma=0;\chi\neq 0;\,\sigma=0$: Configuration tested in \cite{Schwab:2008ce}. 
The parameter $\chi$ would be related to the generation on gravitational field by pressure which is absent in the classical context.
\end{itemize}

In (\ref{dMdrsigma}) we have assumed a quite simple modification of the standard GR mass equation which disappears when $p=p(R)=0$, where $R$ is the star's radius. 
In general, instead of $\sigma p$ one deals with a much complicated function that may depend even on higher derivatives of pressure. 
That dependence is visible if one considers the modified Einstein's equations as
\begin{equation}
 G_{\mu\nu}=T^{\mathrm{eff}}_{\mu\nu}\quad,\quad T^{\mathrm{eff}}_{\mu\nu}=\bar{\sigma}T^{\mathrm{mat}}_{\mu\nu} + W_{\mu\nu}\quad,
\end{equation}
where $T^{\mathrm{mat}}_{\mu\nu}$ is the standard perfect fluid energy-momentum tensor, $\bar{\sigma}$ is a coupling indicating one of the ETG's and $W_{\mu\nu}$ originates from geometric corrections. 
Following that approach one gets that the hydrostatic equilibrium equations differs from (\ref{equilib}) (see for example \cite{santos}).
The TOV equations arising from such equations have much more complicated forms. 
Some of such TOV-like equations can be written as a parametrized TOV equations considered in that work but there is still room for a more detailed 
discussion in this direction (in progress). 
Here, we would like to stress how small modifications of GR stellar equations affect the $M-R$ diagram. 
It is important to point out difficulties appearing in integrating the mass function coming from the new possible modifications. The mass of the NS is calculated as $M=\int_0^R 4\pi r^2 \rho(r) dr$ independent of the functional form involved for the effective mass $\mathcal{M}$ in these equations. 
In GR we have always $\sigma=0$ and therefore there is no difference between the effective mass and the physical meaningful gravitational mass $M$.  
Moreover, there immediately gives rise a question on the stability of the considered system (see {\it Theorem 2} in page 306 of Ref. \cite{weinberg}) which should be also investigated in the terms of ETG's. Here, as in GR, we have adopted the condition $d M / d \rho_c > 0$ for determining the maximum mass.
However, the stability problem for static equilibrium configurations---stars---in ETG is still an open problem in the field.

To close the system of equations an EoS of the type
\begin{equation}
p\equiv p(\rho)\quad,
\label{eq:oS}
\end{equation}
should be specified. 
Equilibirum configurations are found by numerically solving the coupled system (\ref{eq:pTOV}), (\ref{dMdrsigma}) and (\ref{eq:oS}) with the condition on the central pressure $p(0)=p_0$ (corresponding to a central density $\rho_0$) and demanding tha $p(R)=\rho(R)=0$.

\section{Numerical results on the Mass-Radius diagram}\label{secN}

In order to numerically solve the set of equations (\ref{eq:pTOV}), (\ref{dMdrsigma}) and (\ref{eq:oS}) we determine the EoS of the stellar interior.
The usual technique converts $dp/dr$ into $dp/d \rho \, d\rho /dr$ in (\ref{eq:oS}) in such way that the central density ($\rho(r=0) = \rho_0$) becomes a free parameter. 
A given $\rho_0$ value determines one single point in the $M-R$ diagram. 
By varying $\rho_0$ some orders of magnitude around the nuclear saturation density one obtains a curve in this plane $M-R$. 

Among a vast number of possible EoS found in the literature the BSk family provide an unified description of the stellar interior treating in a consistent way  transitions between outer and inner crust (and core). 
The BSk structure has $23$ free coefficients which have to be fitted numerically. 
Each BSk equation of state correspond to one specific numerical fit. For example, the unified BSk19, BSk20, and BSk21 EoSs approximate, respectively, the EoSs FPS \cite{Lorentz, Friedman:1981fj} (soft), APR \cite{Akmal:1998vn} (moderate), and V18 \cite{Li:2008fr} (stiff)---see also \cite{Potekhin:2013si} for discussion and references.

In general, predictions for the maximum masses in BSk models slightly differ. 
BSk19 EoS does not allow masses larger than $2 M_{\odot}$ and therefore observations of very massive neutron stars with radius $\sim 13$~Km would favor the BSk20 and BSk21 fits. 
An up to date compilation of observed neutron stars via different observational methods can be found in \cite{stellarcollapse}. 

\begin{figure}
\begin{center}
\includegraphics[width=1.0\textwidth]{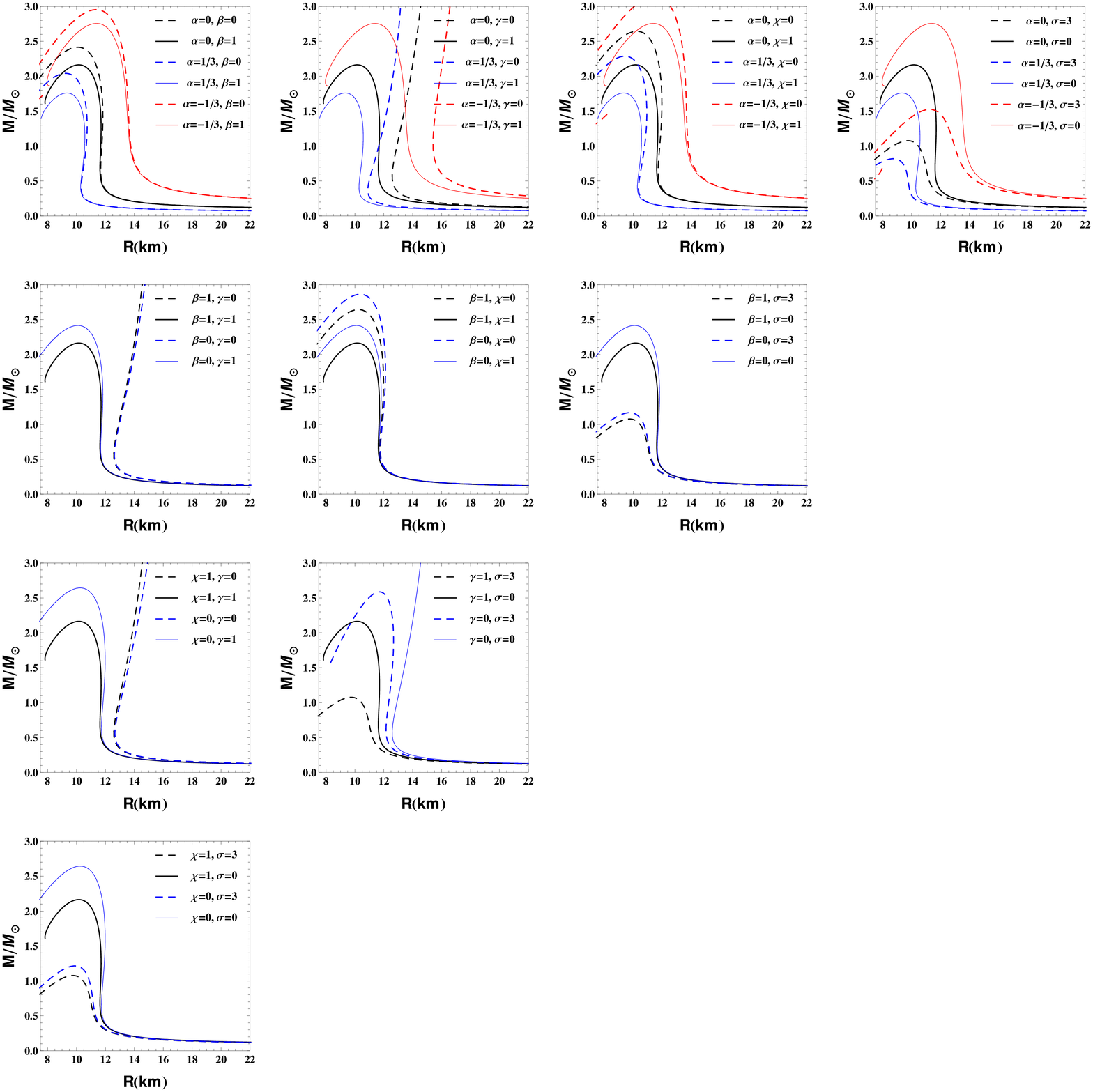}
\caption{Mass-radius diagram for various choices of the parameters $\alpha;\,\beta;\,\gamma;\,\chi;\,\sigma$. In order to mention a specific panel in the text we will introduce the following notation: The panel in the first row (A) and in the first column (I), top-left, is called AI. It this panel the effects of varying $\alpha$ and $\beta$ can be seen while the other parameters are fixed to the GR configuration ($\gamma=1;\,\chi=1;\,\sigma=0$). The same strategy is adopted in the remaining panels. In each panel the standard GR configuration is denoted by the solid-black line. The mass is calculated via ${\rm M}=\int^R_0 4\pi r^2 \rho dr$ even if $\sigma\neq 0$ was adopted when numerically solving the system of equations i.e., we always plot the gravitational mass $M$ rather than the effective mass $\mathcal{M}$. Effects of varying $\sigma$ can be seen in panels $AIV, BIII, CII$ and $DI$.  }
\label{Fig1}
\end{center}
\end{figure}

We will use in this contribution the BSk20 only. 
Our strategy is to fix the GR configuration and let one or two parameters change simultaneously.
This will be shown in Fig. \ref{Fig1}. 
In each panel of this figure the solid black line represents the GR configuration.

\section{Discussion}

In this contribution we have proposed a free `` {\it ad-hoc}'' parametrization of the relativistic equation for equilibrium. 
Our aim is fully pedagogical rather than a proposal for a new extended theory for gravity. By keeping the BSk20 EoS for the neutron star interior--- which is realistic in the sense that equilibrium configurations achieve $2 M_{\odot}$ using GR---we seek the impact of the new parameters $\alpha, \beta, \gamma, \chi$ and $\sigma$. 
They can be visualized in Eq. (\ref{eq:pTOV}). 

The simplest expected manifestation of modified gravity theories occurs via a redefinition of the effective gravitational coupling $G_{eff}\rightarrow G(1+\alpha)$.  
The effects of $\alpha$ on the $M-R$ plane are seen in the panels belonging to row $A$ (see the caption of Fig. \ref{Fig1}). 
The larger the $\alpha$ value, the smaller the typical radius of the equilibrium configurations. 
The maximum mass is also reduced. 

The parameter $\beta$ is clearly related to the maximum mass. The larger the inertial effects of pressure ($\beta$), the smaller the maximum mass.   

Concerning the parameter $\chi$ we have confirmed the findings of \cite{Schwab:2008ce}, i.e., the self-gravity of pressure contributes to reducing the maximum allowed NS with almost no impact on the radius of the star.

The impact of parameter $\gamma$ is relevant to panels $AIII, BI$ and $CI$. In the Newtonian hydrostatic equilibrium $\gamma=0$. However, 
by re-performing the Newtonian classical case with a modified Newtonian potential of the type $V(r)=-\frac{GM}{r-A} $, where $A$ is a constant, it 
is possible to obtain a similar contribution with a singularity in the denominator of the TOV equation like the $\gamma=1$ case. 
The existence of a maximum mass (absent in the Newtonian case) is a pure relativistic prediction. But it is worth noting that the curves with $\gamma=0$ produce indeed a maximum mass, but with a very large value which can not be seen in these plots.



\end{document}